\documentclass{article} 
\begin{document}

\title{The quasars are neutron stars embedded in clouds of hydrogen.}

\author{ Jacques Moret-Bailly \footnote{Address: 265,rue St Jean F-21850 St Apollinaire, France.
e-mail: Jacques.Moret-Bailly@u-bourgogne.fr}
}

 \maketitle

 \begin{abstract}
The accretion of a cloud of hydrogen at the surface of a small, heavy star produces a high energy mostly dissipated 
by electromagnetic radiation.
The combination of the absorption and the redshift of this radiation by hydrogen explains all spectral observations: 
shapes of the broad lines, and their anti-correlation with the radio-loudness, lack of lines in front of the Lyman forest, 
value $z_b = 0.062$ of the periodicity of the redshifts, correlation of a high redshift with a thermal spectrum attributed 
to dust.
These properties are deduced from usual laws of physics without new parameters, they do not require any non 
baryonic matter.

11.17.3; (Galaxies:) quasars: general.

\end{abstract}

 \section{Introduction}
There are many reasons to question the current explanations for the spectra of quasars.  According to theory, to 
produce the Lyman lines in the redshift of quasars, atomic hydrogen is concentrated into bands or islands at many 
different redshifts; or conversely relativistic jets flow from the faces of quasars. Both hypothesis are flawed because 
they require clouds of hot atomic hydrogen, whose stability and/or velocity cannot be explained using conventional 
physical models.

To understand propagation of light in low pressure gases, one must take into account the  \textquotedblleft Coherent 
Raman Effect on time-Incoherent Light\textquotedblright  (CREIL), which transfers energy by frequency shifts from 
shorter wavelengths to longer wavelengths without  blurring or distorting the images or the spectra. The relative 
frequency shifts  $\Delta\nu/\nu$  due to CREIL are nearly constant.

In a previous paper  (Moret-Bailly \cite{Mor03}), we explained that the spectra can be obtained by assuming that a 
nearly homogeneous cloud of Lyman pumped atomic hydrogen is perturbed by a variable magnetic field: Where the 
field is low, there is virtually no redshift and the absorption (or emission) lines are written visibly into the spectrum. If 
the field is high, the redshift is simultaneous with the absorption (or emission),  blurring the lines and making them 
invisible. The spectrum results not from a modulation of the absorption, but from the redshifting function of the gas. 
We also showed that the existence of a non-linearity is able to increase the contrast of the lines.

However our previous explanation requires an important variation of the magnetic field for each line, which in turn 
requires the existence of a large number of magnetised satellites. Another weakness in this argument is that the 
spectra show a pseudo-stochastic repartition of the redshifts of the lines, the difference of two redshifts being the 
product of a constant  $z_b  = 0.062$  multiplied by an integer (Burbidge \cite{Burbidge}, Tifft \cite{Tifft76,Tifft95}, 
Burbidge \& Hewitt  \cite{Hewitt}, Bell \cite{Bell}, Bell \& Comeau \cite{Comeau}). It seems very difficult to explain 
such a spectroscopic regularity of the \textquotedblleft intrinsic redshift\textquotedblright by the presence of objects 
in the line of sight.  The non-linearity we introduce in this letter provides a very simple solution.

\medskip
Section \ref{CREIL} recounts the key spectroscopic properties which provide the \textquotedblleft Coherent Raman 
Effect on time-Incoherent Light\textquotedblright (CREIL) able to explain the intrinsic redshifts. 

Section  \ref{H}  applies CREIL to a halo of atomic hydrogen; to obtain the required resonances, H must have a non-
zero orbital quantum number. Lyman pumping is demonstrated  to provide the necessary quantum state and  non-
linearity.

Section \ref{tree} explains the appearance of the periodic redshifts.

Section \ref{neutron} explains the building of the spectrum of a quasar during the propagation of the light in the 
cloud of hydrogen which surrounds it.

\section{The  \textquotedblleft Coherent Raman Effect on time-Incoherent Light\textquotedblright 
(CREIL)}\label{CREIL}
The CREIL is very similar to the \textquotedblleft Impulsive Stimulated Raman Scattering\textquotedblright (ISRS)  
discovered in 1968  (Giordmaine \cite{Giordmaine})  and commonly used to study radiation mechanics (Yan et al. 
\cite{Yan},  Weiner et al. \cite{Weiner}, Dougherty et al. \cite{Dougherty}, Dhar et al. \cite{Dhar}).

Both effects are space-coherent, they do not blur the images. They are parametric, using matter as a 'catalyst', leaving 
the molecules unchanged. They may be considered as a combination of two nearly simultaneous coherent Raman 
scatterings yielding virtually opposite transitions. The relaxation times of the matter used as a catalyst must be longer 
that the length of the light pulses; these relaxation times are the period of (at  least) a quadrupole-allowed transition, 
and, in a gas, the collisional time.

The difference results from the use of ordinary time-incoherent light in the CREIL while the ISRS uses strong 
ultrashort laser pulses. In the ISRS, the strong laser flux has a qualitative effect: the Raman scattered amplitude which 
produces the frequency shifts (by interference with the exciting light beam), increases as a nearly quadratic function 
of the exciting flux amplitude. In the CREIL, at low power, all scatterings are linear: under the described conditions the 
coherently scattered amplitude is proportional to the exciting amplitude.  This behavior, makes the frequency shift 
independant of the intensity.

The ISRS and the CREIL may be computed from the tensors of polarisability of the molecules. This is only slightly 
dependent on the exciting frequencies, so that, in a first approximation, the relative frequency shift $\Delta\nu/\nu$  
does not depend on the exciting frequency; as the the redshifts produced by the CREIL do not depend on the 
intensities, the spectra are slightly distorted, but not blurred.

The length of the pulses of ordinary (thermal) incoherent light is of the order of 5 nanoseconds. To avoid an 
incoherent scattering, the mean molecular collision rate must be greater than 5 ns. This is only possible when the 
pressure is less than a few pascals. To produce a CREIL effect, the Raman resonance period must also be longer than 
5 ns, that is a Raman resonance frequency  in the MHz range (Moret-Bailly \cite{Mor98a,Mor98b,Mor01}). Both of 
these conditions increase the path length necessary for a measurable frequency shift. Because of these extremes, 
CREIL has never been measured in a laboratory. In space the pressure is generally low, and the light paths are long. 
The 2.7K background thermal radiation, provides a uniform low temperature exciting energy that is blueshifted 
(heated) during a CREIL event.  

The other essential component in a CREIL event is a molecular catalyst with a low frequency quadrupolar resonance. 
Hydrogen is the most common element and exists in several allotropic states, mono-, di- and tri-atomic. Some 
polyatomic allotropes (H$_2^+$, H$_2$D$^+$, ...) exhibit Raman hyperfine resonances close to 30 Mhz; although 
these molecules are not readily observed, they undoubtedly occur due to ultraviolet radiation and collisions. The 
reason they are not observed  is very simple: these molecules are destroyed by the collisions with H$_2$ so that their 
half-life is long only where the pressure is low enough to allow CREIL. A simultaneous absorption and CREIL makes 
absorption lines as wide as the redshift; therefore the lines of these molecules are so wide and weak that they cannot 
be seen. Atomic hydrogen also provides convenient transitions if it has been pumped by a Lyman transition so that 
hyperfine structures appear in non-zero quantum shells.

\section{Propagation of light in atomic hydrogen.} \label{H}

Consider the propagation of light having a continuous spectrum (constant intensity, in particular in the  Lyman 
region), in an homogeneous atmosphere of low pressure atomic hydrogen. In the fundamental state  (principal 
quantum number n=1), the distance between the hyperfine levels (1420 MHz) is too large. In the  other states, 
hyperfine transitions have convenient frequencies for the Raman allowed selection rule $\Delta  F = 1$, for instance : 
178 MHz in $2s_{1/2}$, 59 MHz in $2p_{1/2}$ and 24 MHz in $2p_{3/2}$.

Set $\Delta L$ the length of path for which the redshift is equal to the linewidth $\delta\nu$ of the Lyman  $\alpha$ 
line, and assume that the atoms which are active in CREIL are mostly pumped by the Lyman  $\alpha$ transition.

Set $\Delta\nu$ the redshift along the path $\Delta L$, which {\it would} result from a {\it complete}  Lyman $\alpha$ 
absorption of the intensity $I$ , and suppose that, in a first approximation, the whole  redshift results from the Lyman 
$\alpha$ absorption.

- case a: If $\Delta\nu$ is larger than the Lyman $\alpha$ linewidth $\delta\nu$ , that is if $I$ is large enough, $I$ is not 
fully absorbed, only the {\it constant} intensity $\Delta I$ which produces the redshift  $\delta\nu$ is subtracted from 
$I$, that is from the spectrum while, by redshifting, the Lyman line crosses  it. Thus, the contrast of lines which have 
been written into the spectrum is increased.

- case b: If, on the contrary, $\Delta\nu$ is lower than $\delta\nu$, the first approximation fails, a part of  the redshift 
must result from other Lyman absorptions or other active atoms. Assuming a low redshifting  power for these effects, 
a long path $\Delta L$ is necessary to get the redshift $\delta\nu$, so that the  absorption of all lines is strong.

\medskip

If the intensity $I$ is constant and high, except for a single absorption line, the redshift and the absorption  are 
constant (case a), except at a coincidence of the line with a Lyman line; at this coincidence, the  redshifting power 
decreases (strongly if case b is reached), so that the absorption of {\bf all lines} of the  gas is increased; similarly, a 
written emission line increases the redshifting power, so that the decrease of  absorption appears as an emission; {\it 
the coincidence by redshift of a line already written in the spectrum  with a Lyman line writes the whole spectral 
pattern of the gas into the spectrum.}

\section{Building the periodic redshifts.}\label{tree}

Suppose that a single Lyman pattern is written in the spectrum. The coincidence of the written, redshifted  Lyman 
$\beta$ (resp. Lyman $\gamma$) line with the Lyman $\alpha$ line of the gas writes the Lyman  pattern into the gas. 
Both patterns differ by the shift of frequencies $\nu_{(\beta \,{\rm resp.}\, \gamma)}- \nu_\alpha$ of the $\alpha$ and 
$\beta$ \,(resp. \,$\gamma$) lines. As in the standard computations the  lines are considered as Lyman $\alpha$, the 
frequency shift is relative to the Lyman $\alpha$ frequency:

\begin{equation}
\begin{array}{l}
\qquad z_{(\beta \,{\rm resp. }\,\gamma) , \alpha}=\frac{ \nu_{(\beta \,{\rm resp.} \,\gamma)}-\nu_\alpha}{  
\nu_\alpha} \approx \\
\approx\frac{1-1/(3^2 \,{\rm resp. \,}4^2)-(1-1/2^2)}{1-1/2^2}
\end{array}
\end{equation}

\begin{equation}
\begin{array}{l}
z_{(\beta , \alpha)}\approx 5/27 \approx 0.1852 \approx 3*0.0617; \\  z_{(\gamma , \alpha)}= 1/4 = 0.25 =4*0.0625.
\end{array}
\end{equation}

Similarly $ z_{(\gamma , \beta)}\approx 7/108 \approx 0.065.$ The redshifts appear, with a good approximation as the 
products of $z_b = 0.062$ by an integer $q$.

The intensities of the Lyman lines are decreasing functions of the final principal quantum number $n$, so  that the 
inscription of a pattern is better for $q =3$ than for $q = 4$ and {\it a fortiori} for $q = 1$.

\medskip Iterating, the coincidences of the shifted line frequencies with the Lyman $\beta$ or $\alpha$ frequencies  
build a tree, final values of $q$ being sums of the basic values 4, 3 and 1. Each step being characterised by  the value 
of q, a generation of successive lines is characterised by successive values of $q: q_1, q_2...$ As  the final redshift is 
$q_F*z_b = (q_1 +q_2 +...)*z_b $, the addition $q_F = q_1 + q_2 +...$ is both a  symbolic representation of the 
successive elementary processes, and the result of these processes.

 The name \textquotedblleft tree\textquotedblright\, is not very good because 'branches' of the tree may be 'stacked' 
by coincidences of frequencies. A remarkable coincidence happens for $q = 10$, this number  being obtained by the 
effective coincidences deduced from:

\begin{equation} 10=3+3+4=3+4+3=4+3+3=3+3+3+1=... \end{equation} 
$q = 10$ is so remarkable that $z_f = 10z_b = 0.62$ may seem experimentally a value of $z$ more  fundamental than 
$z_b$.

 \medskip
In these computations, the levels for a value of the principal quantum number $n$ larger than 4 are neglected by 
assuming that the corresponding transitions are too weak.

\section{The model of quasar.}\label{neutron}

The accretion of hydrogen produces a high energy, thus a high temperature at, and close to the surface of the quasar. 
As this region is not a black body, the intensities of the emission lines is larger than the background intensities.

Close to the quasar, the atomic hydrogen if fully ionised, it does not produce redshifts, metal lines may be relatively 
sharp.

{\it The broad lines:}

 If the star emits a strong electromagnetic field (radio-loud quasars) the ions are accelerated, so that the temperature 
remains high over a large distance, the hydrogen remains ionised until a pressure of an order of magnitude of a Pascal 
increases the collisional time enough to break the heating and ionising process.

 If the star is radio-quiet, some neutral hydrogen remains at pressures larger than a Pascal: Lyman emission, then 
absorption lines appear. These lines are saturated, thus broad; their intensity corresponds, in a large part of their 
width, to an equilibrium of the temperatures of the gas and of the light. The existence of excited neutral atoms induces 
the redshift process described previously.

{\it The \textquotedblleft proximity effect\textquotedblright:}

Between the emission and absorption regions, the gas has approximately the temperature of the light, so that no lines 
are written; however some atoms are excited, so that there is a region without visible lines over a redshift of the order 
of z=0.5: it is the proximity effect (Rauch \cite{Rauch}).

{\it The Lyman forest and the \textquotedblleft dust\textquotedblright:}

The Lyman forest corresponds to the process described in section \ref{tree}. If the redshift of the emission lines is 
large, a lot of energy is transferred to the thermal radiation; it seems that this thermal radiation is produced by hot 
dust.

{\it The high redshifts: }

The \textquotedblleft intrinsic redshift\textquotedblright produced by the cloud which surrounds the quasar is much 
larger than the remaining redshift for which Hubble's law is valuable; thus the quasar is not very far, it may be simply 
a neutron star.

\section{Conclusion.}

The present computation is a quantitative explanation of the frequencies observed in the spectra of the quasars. It 
requires only elementary physics, no unknown matter. The computation should be improved by an evaluation of the 
intensities.

The CREIL appears as a key in the study of the quasars. It should probably be helpful for other studies:  for instance, 
some astrophysicists think that the dark matter needed to get the gravitational stability could  be simply molecular 
hydrogen; with this hypothesis, it exists, by UV ionisation, some H$_2^+$ which, by collisions with H$_2$, generates 
H$_3$ and H$_3^+$; the hyperfine structures of H$_2^+$, H$_2D^+$ and H$_2D$, the Rydberg levels of H$_3$ 
provide quadrupolar resonances in the megahertz range, therefore a contribution of the CREIL not only to the 
\textquotedblleft intrinsic\textquotedblright redshifts, but to the \textquotedblleft cosmological\textquotedblright 
redshifts. A convenient, constant density of excited hydrogen in the intergalactic space gives Hubble's law, without 
expansion of the Universe.

\section{acknowledgements.}

I thank very much Jerry Jensen who improved and corrected the manuscript, and Iain. R. McNab for a discussion on 
the allotropes of hydrogen.

\end{document}